\documentclass{XrU2005}

\usepackage[]{natbib}

\usepackage{graphicx}

\usepackage{amssymb}

\def \arvind    {A.~N. Parmar}
\def \laurence  {L. Boirin}
\def \mariano {M. M{\'e}ndez}
\def \maria {M. D{\'i}az Trigo}
\def \diaz {D{\'i}az Trigo}
\def \jelle {J.~S. Kaastra}

\def \sron {The SRON, the National Institute for Space Research,
  Sorbonnelaan 2, 3584 CA Utrecht, The Netherlands}

\def \strasbourg {Observatoire Astronomique de Strasbourg, 11 rue de
l'Universit\'e, F-67000 Strasbourg, France}

\def \estec {Research and Scientific Support Department of ESA, ESTEC,
                Postbus 299, NL-2200 AG Noordwijk, The Netherlands}

\def \rsun {\ifmmode$R$_{\odot}\else R$_{\odot}$}

\def \hcm {\hbox {\ifmmode $ atoms cm$^{-2}\else atoms cm$^{-2}$\fi}}

\def\approxgt{\mathrel{\hbox{\rlap{\lower.55ex \hbox {$\sim$}}
        \kern-.3em \raise.4ex \hbox{$>$}}}}
\def\approxlt{\mathrel{\hbox{\rlap{\lower.55ex \hbox {$\sim$}}
        \kern-.3em \raise.4ex \hbox{$<$}}}}

\def \nh {$N{\rm _H}$}
\def \nhabs {$N{\rm _H^{abs}}$}
\def \nhxabs {$N{\rm _H^{xabs}}$}

\newcommand {\fetfive} {\ion{Fe}{xxv}}
\newcommand {\fetsix} {\ion{Fe}{xxvi}}

\def \xiunit {\hbox{erg cm s$^{-1}$}}
\def \logxi {$\log(\xi)$}

\def \xabs {{\tt xabs}}

\def \kgau {$k_{\rm gau}$}

\def \nineteen {XB\,1916$-$053}
\def \mxb {MXB\,1658$-$298}
\def \bigdip {X\,1624$-$490}
\def \twelve {X\,1254$-$690}
\def \grs {GRS\,1915+105}
\def \gro {GRO\,J1655$-$40}
\def \gx {GX\,13+1}
\def \cir {Cir\,X$-$1}
\def \exo {EXO\,0748$-$676}

\def \thirteen {4U\,1323$-$62}
\def \seventeen {H\,1743$-$322}

\def \stfs {4U\,1746$-$371}

\def \src {\thirteen}

\def \xmm {XMM-Newton}

\DeclareRobustCommand{\ion}[2]{%
\relax\ifmmode
\ifx\testbx\f@series
{\mathbf{#1\,\mathsc{#2}}}\else
{\mathrm{#1\,\mathsc{#2}}}\fi
\else\textup{#1\,{\mdseries\textsc{#2}}}%
\fi}

\def\aap{A\&A}%
\title{A highly-ionized absorber as a new explanation for the spectral
changes during dips from X-ray binaries}

\author{\laurence}
\affil{\strasbourg}

\author{\maria}
\affil{\estec}

\author{\mariano}
\affil{\sron}

\author[2]{\arvind}
\author[3]{\jelle}

\begin{document}


\maketitle

\begin{abstract}
  
  Until now, the spectral changes observed from persistent to dipping
  intervals in dipping low-mass X-ray binaries were explained by
  invoking progressive and partial covering of an extended emission
  region. Here, we propose a novel and simpler way to explain these
  spectral changes, which does not require any partial covering and
  hence any extended corona, and further has the advantage of
  explaining self-consistently the spectral changes in both the
  continuum and the narrow absorption lines that are now revealed by
  XMM-Newton. In \src, we detect \fetfive\ and \fetsix\ absorption
  lines and model them for the first time by including a complete
  photo-ionized absorber model rather than individual Gaussian
  profiles.  We demonstrate that the spectral changes both in the
  continuum and the lines can be simply modeled by variations in the
  properties of the ionized absorber. From persistent to dipping the
  photo-ionization parameter decreases while the equivalent hydrogen
  column density of the ionized absorber increases.  In a recent work
  (see \diaz\ et al. in these proceedings), we show that our new
  approach can be successfully applied to all the other dipping
  sources that have been observed by XMM-Newton.

\end{abstract}

\section{Introduction}

\begin{figure*}[!t]
\centerline{\hspace{-1cm}\includegraphics[angle=90,width=1.15\textwidth]{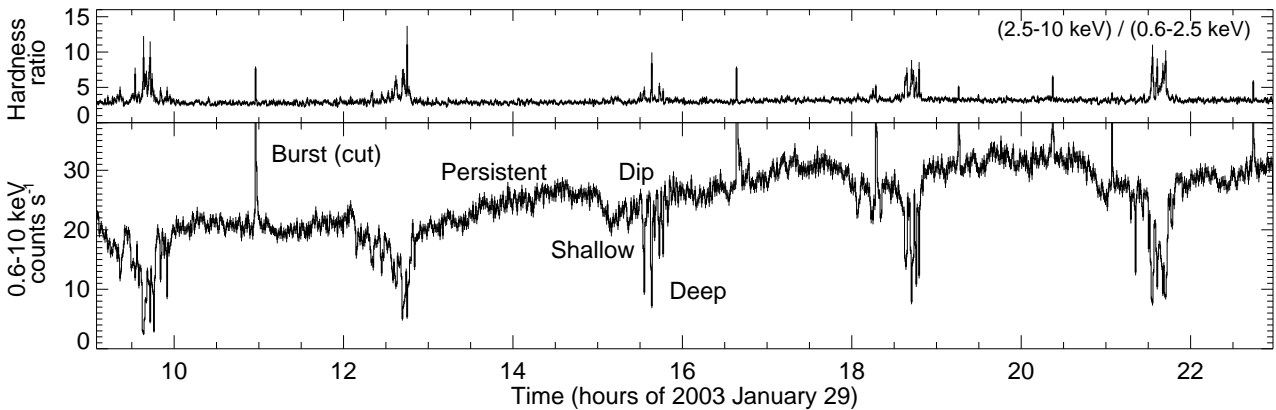}}
{\vspace{-1.5cm}}
\caption{0.6--10~keV EPIC PN lightcurve of \src\ ({\bfseries
    bottom}). The dipping activity is associated with spectral
  hardening ({\bfseries top)}. Adapted from \cite{1323:boirin05aa}.}
\label{fig:lc}
\end{figure*}

The lightcurves from dipping low-mass X-ray binaries (LMXBs) such as
\src\ show dips recurring at the orbital period of the system
(Fig.~\ref{fig:lc} bottom). Dips are due to a structure passing
through the line-of-sight at each orbital rotation. This structure is
probably a thickened region of the disk related to the impact of the
stream from the companion star into the disk.  The presence of
periodic dips and absence of eclipses from the companion indicate that
dipping sources are viewed relatively close to edge-on.

The X-ray spectra of most of the dip sources become harder during
dipping (Fig.~\ref{fig:lc} top).  However, simple photo-electric
absorption by cool (neutral) material fails to explain the spectral
changes from persistent to dipping intervals. Therefore, more complex
models have been proposed. In particular, the ``complex continuum''
approach has been successfully applied to a number of dipping LMXBs
including \src\ \citep[][]{1323:balucinska99aa}.  It assumes that the
X-ray emission originates from two components, and the spectral
changes during dips are explained by the partial and progressive
covering of one of the components by a cool absorber, while the other
component is rapidly and entirely covered by another cool
absorber. This approach implies that the latter component comes from a
point-like region such as the neutron star surface, whereas the former
component comes from a very extended corona.

%
%

The improved sensitivity and spectral resolution of {\it Chandra} and
XMM are allowing narrow absorption features from highly ionized Fe and
other metals to be observed in a growing number of X-ray binaries.  In
particular, \fetfive\ (He-like) or \fetsix\ (H-like) resonant 1s-2p
absorption lines near 7~keV were reported from the micro-quasars \gro,
\grs\ and \seventeen, and from the neutron star systems \cir, \gx,
\mxb, \bigdip, \twelve, \nineteen\ and now \thirteen\
\citep[references in][]{1323:boirin05aa}. These sources are known to
be viewed close to edge-on (many are dippers).  This indicates that
the highly ionized plasma probably originates in an accretion disk
atmosphere or wind, which could then be a common feature of accreting
binaries but preferentially detected in systems viewed close from the
disk plane.

Here, we report the detection of \fetfive\ and \fetsix\ absorption
lines from the LMXB \src\ and propose a new explanation for the
spectral changes between persistent and dipping intervals
\citep[details in ][]{1323:boirin05aa}. We further show that this new
explanation also applies to all the other bright dipping sources
observed by \xmm\ \citep[details in][]{diaztrigo05aa}.

\section{Results on \src}

We analyzed the 50~ks XMM-Newton observation of \src\ performed on
2003 January 29 (Fig.~\ref{fig:lc}). Bursts were excluded and one
spectrum was extracted for each category of emission: persistent,
shallow dipping and deep dipping. \fetfive\ and \fetsix\ 1s-2p
resonant absorption lines near 7~keV are clearly detected in the
persistent spectrum (Fig.~\ref{fig:1323}~A top), indicating that a
highly-ionized disk atmosphere or wind is present in \src. Absorption
lines are also present in the dipping spectra (Fig.~\ref{fig:1323}~A
middle and bottom) indicating that the structure causing the dips
(``bulge'' hereafter) is also ionized. However, clear spectral changes
in the lines are visible from persistent to deep dipping: the strength
of the \fetsix\ line decreases while that of \fetfive\ increases,
indicating that the bulge is less strongly ionized.

For the first time, to account for the absorption features evident
near 7~keV, we include a photo-ionized absorber in the spectral model,
rather than individual line profiles.  We use the \xabs\ model of
SPEX, which treats the absorption by a thin slab composed of different
ions, located between the ionizing source and the observer.  The
processes considered are the continuum and the line absorption by the
ions and scattering out of the line-of-sight by the free electrons in
the slab.  The relevant ions are automatically taken into account and
their relative column densities are coupled in a physical way via a
photo-ionization model.

We find that the persistent and dipping spectra are all well fit by a
model consisting of a power-law, a blackbody and a broad Gaussian
emission line, modified by absorption from neutral ({\tt abs}) and
ionized ({\tt xabs}) material (Fig.~\ref{fig:1323}~B and C).  The
ionized plasma has a lower ionization parameter and a larger column
density during dipping.  In all cases, it perfectly accounts for the
narrow features near 7~keV.  Remarkably, it also produces apparent
continuum absorption which becomes substantial and strongly
energy-dependent during dipping (compare panels~d in
Fig.~\ref{fig:1323}~B and C).  Indeed, because the ionization is lower
during dipping, there is a wider variety of ions than during
persistent emission where most of the species are fully stripped of
their electrons. Thus many more absorption lines and edges are
expected during dipping (see Fig.~\ref{fig:trans}~A).  Furthermore,
because the column density is larger, the edges are stronger.  This
explains the smooth variation of the transmission with energy (outside
the sharp changes at the binding energies themselves).

%
%
\begin{table}[!b] \caption{The column density of the neutral (\nhabs) and
ionized (\nhxabs) absorbers and the ionization parameter \logxi\ in
\src. During dipping, the ionized absorber has a lower ionization
level and a larger column density. There is also more neutral absorber.}
\begin{center}
\begin{tabular}{llccc}
\hline
\noalign {\smallskip}
\multicolumn{2}{l}{}  & Persistent & Shallow dip & Deep dip\\
\noalign {\smallskip}
\multicolumn{2}{l}{\nhabs\ }&   $3.50 \pm 0.02$             & $3.58 \pm 0.03$       &  $4.2 \pm 0.2$ \\
\multicolumn{2}{l}{ \nhxabs\ } &   $3.8 \pm 0.4$       &  $14 \pm 1$   & $ 37 \pm 2$   \\
\multicolumn{2}{l}{\logxi\ }   &  $3.9 \pm 0.1$      &  $3.43 \pm 0.08$      &  $3.13 \pm 0.07$ \\
\noalign {\smallskip}
\hline
\noalign {\smallskip}
\multicolumn{5}{l}{\small \nhabs\ and \nhxabs\ are in $10^{22}$ cm$^{-2}$ and $\xi$ in
\xiunit.}
\end{tabular}
\end{center}
\label{tab:results}
\end{table}

By successfully fitting the dipping spectra using the persistent
model, but fixing the parameters of the continuum to the persistent
values, and allowing only the parameters of the absorbers ({\tt abs}
and {\tt xabs}) to change, we actually demonstrate that the spectral
changes from persistent to dipping can be modeled simply by variations
in the properties of the neutral and ionized absorbers, with the
ionized absorber playing the main role (Table~\ref{tab:results}).
Contrary to the ``complex continuum'' model, the new proposed approach
does not require any partial covering and hence does not require the
underlying source of X-ray emission to be particularly extended in
\src. The new explanation further presents the advantage of explaining
self-consistently the spectral changes both in the continuum
\emph{and} the narrow lines.

\begin{onecolumn}

%
%
\begin{figure*}[th!]
\centerline{
\includegraphics[angle=0,width=0.33\textwidth]{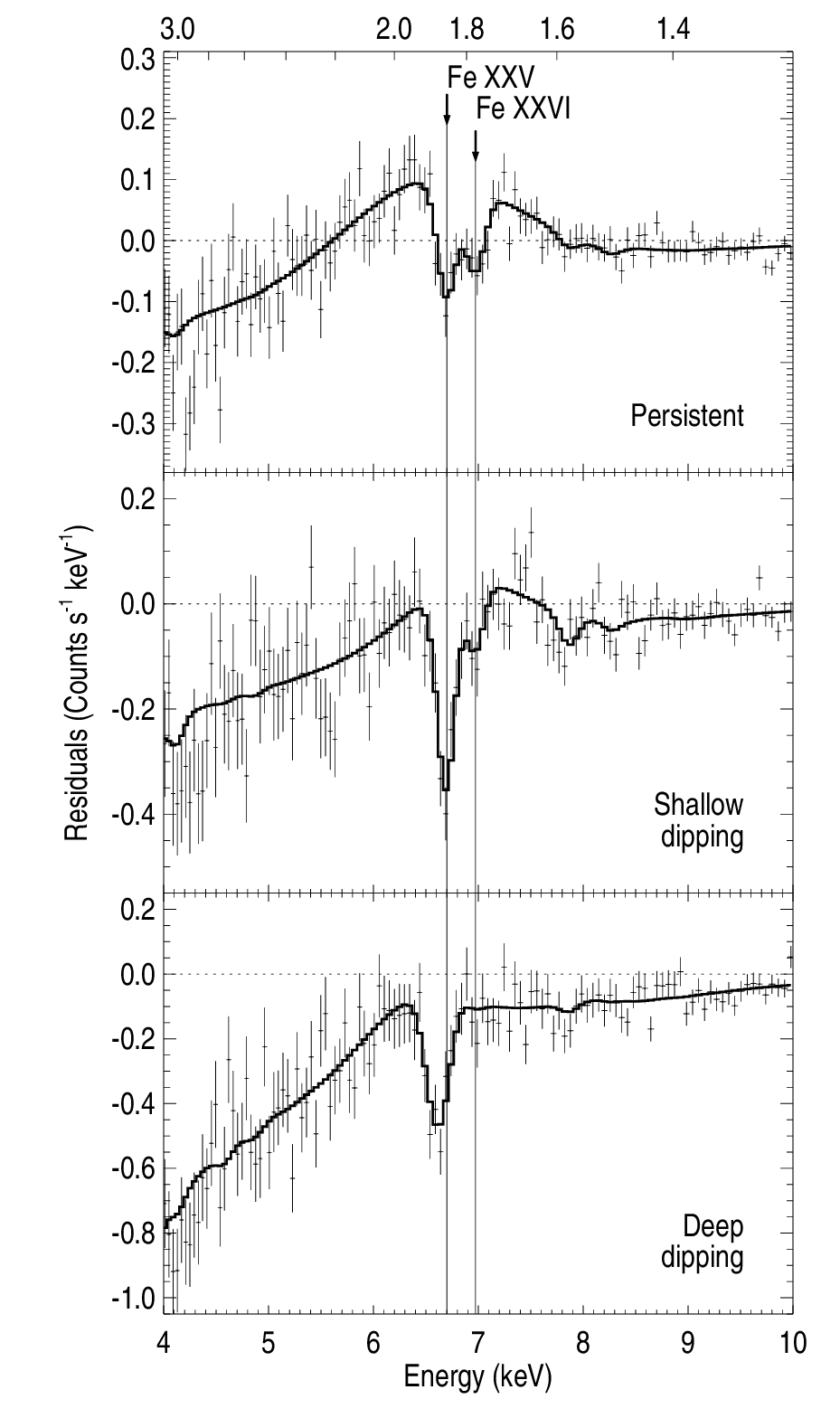}
\includegraphics[width=0.33\textwidth]{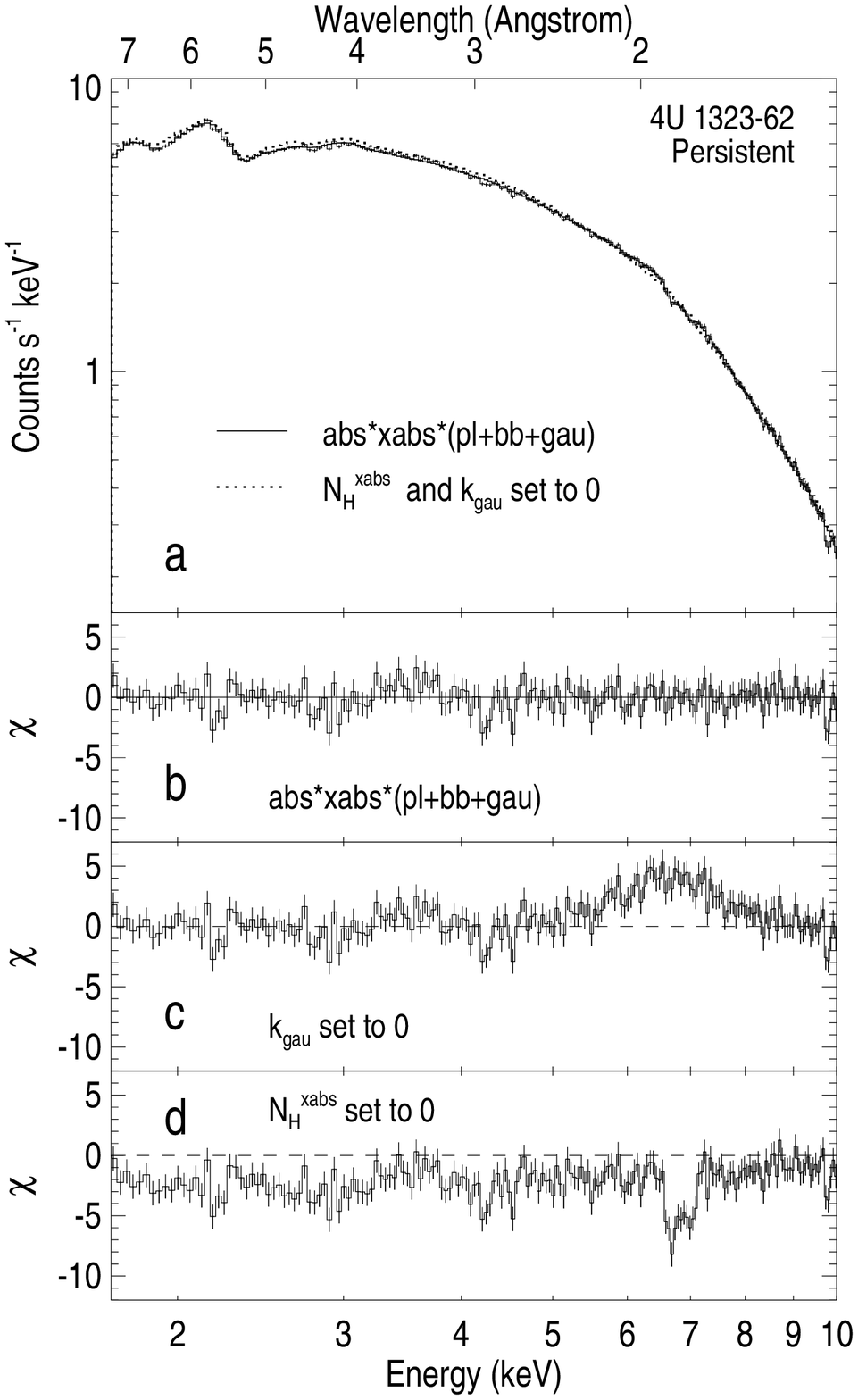}
\includegraphics[width=0.33\textwidth]{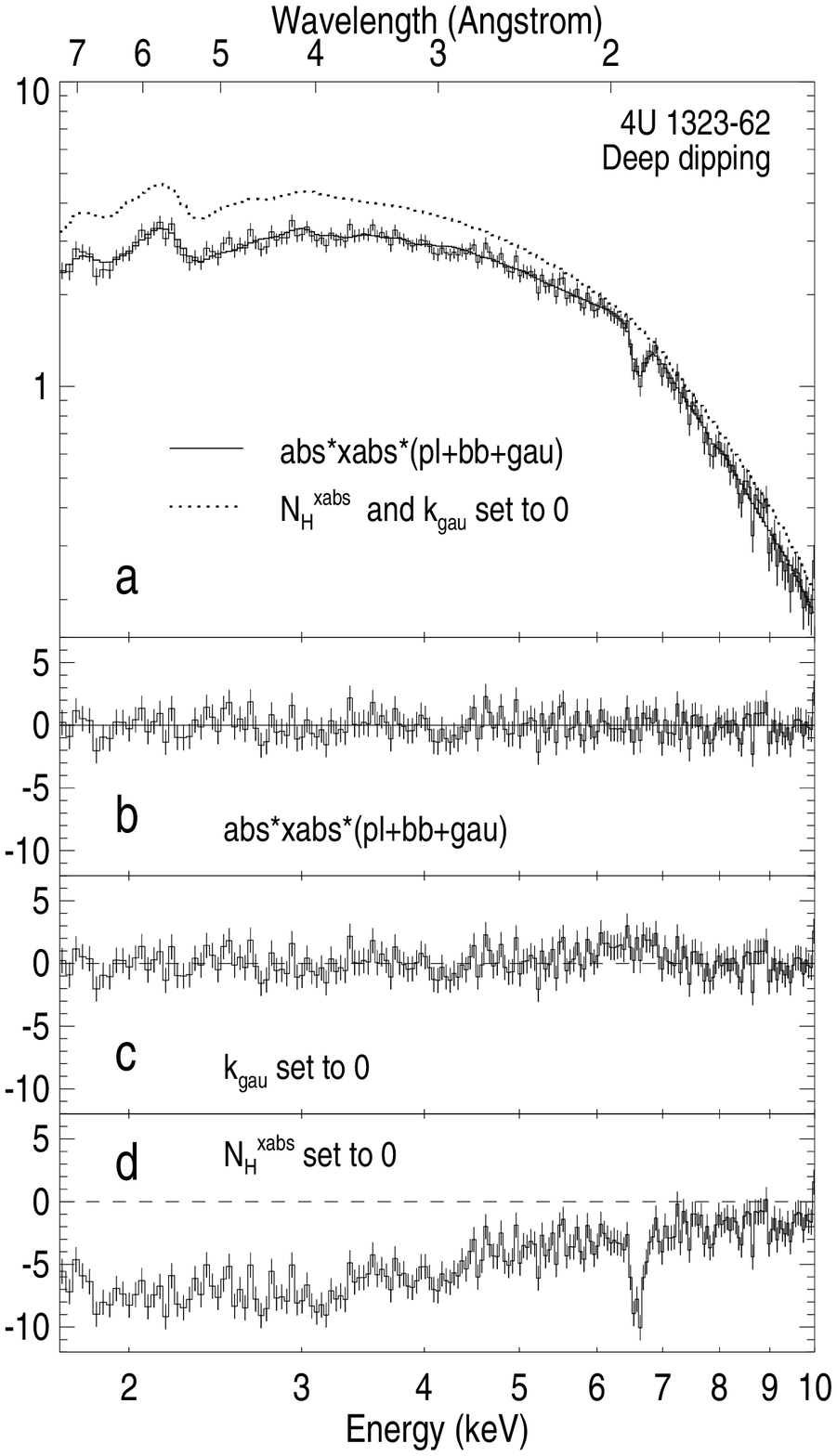}
}
{\vspace{-0.8cm}{\bfseries A \hfill B \hfill C \hfill}}
\caption{EPIC PN results on \src\ from
  \citet{1323:boirin05aa}. {\bfseries A)} 4--10 keV spectral residuals
  showing the \fetfive\ and \fetsix\ absorption lines during
  persistent ({\bfseries top}) and shallow ({\bfseries middle})
  emission.  During deep dipping ({\bfseries bottom}), the \fetsix\
  line is not present anymore: the absorber is less strongly ionized.
  {\bfseries B)} {\bfseries a)} Persistent spectrum fit with a model
  consisting of a power-law ({\tt pl}), a blackbody ({\tt bb}) and a
  broad Gaussian emission line ({\tt gau}), modified by absorption
  from neutral ({\tt abs}) and ionized ({\tt xabs})
  material. {\bfseries b)} Flat residuals from the above model
  indicating that the fit is good.  {\bfseries c)} Residuals showing
  the contribution of the Gaussian emission line at 6.6~keV (by
  setting its normalization,\kgau, to 0).  {\bfseries d)} Residuals
  showing the contribution of the ionized absorber (by setting
  \nhxabs\ to 0). It perfectly accounts for the narrow \fetfive\ and
  \fetsix\ absorption lines.  {\bfseries C)} Same as {\bfseries B} but
  for the deep dipping intervals. The ionized absorber does not only
  produce the line near 7~keV, but also energy-dependent absorption
  throughout the spectrum (panel {\bfseries d}).}
\label{fig:1323}
\end{figure*}

%
%
\begin{figure*}[bh!]
\centerline{\hspace{-0.4cm}\includegraphics[origin=c,angle=90,width=0.5\textwidth]{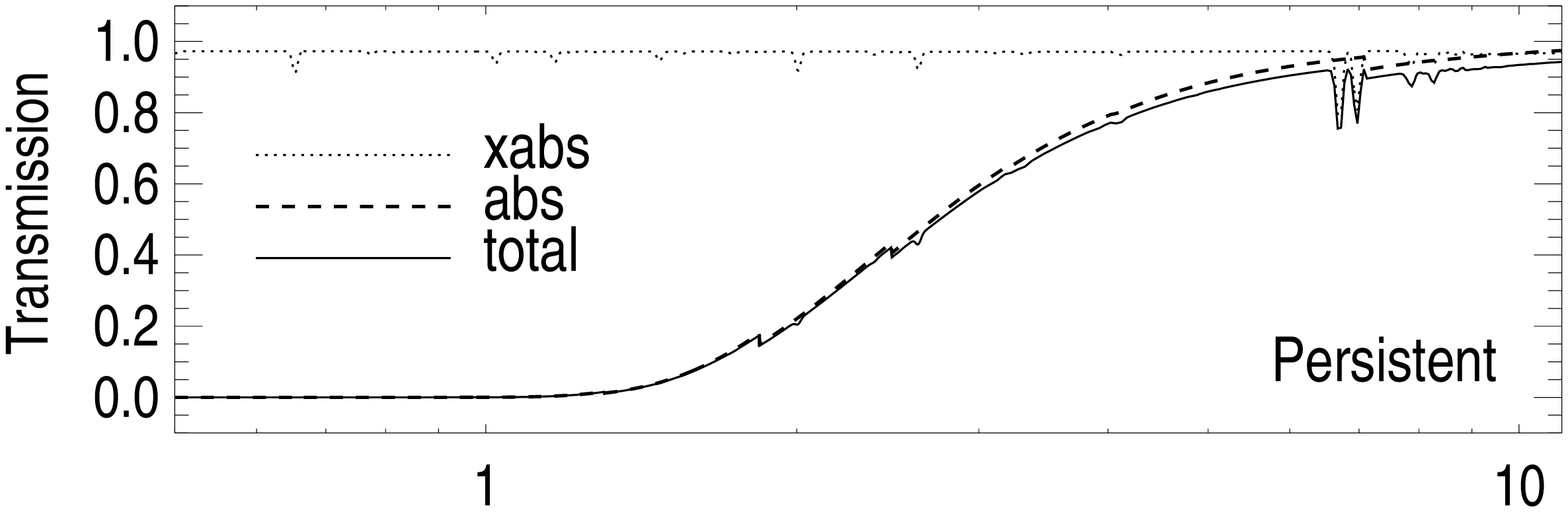}\hfill
\includegraphics[origin=c,angle=-90,width=0.5\textwidth]{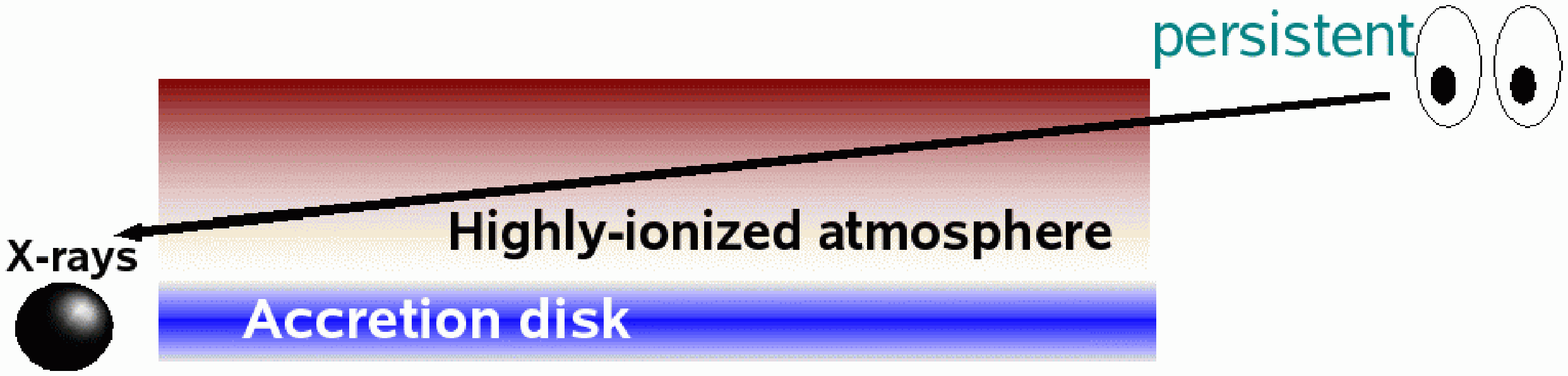}}
\vspace{-3.35cm}
\centerline{\hspace{-0.4cm}\includegraphics[origin=c,angle=90,width=0.5\textwidth]{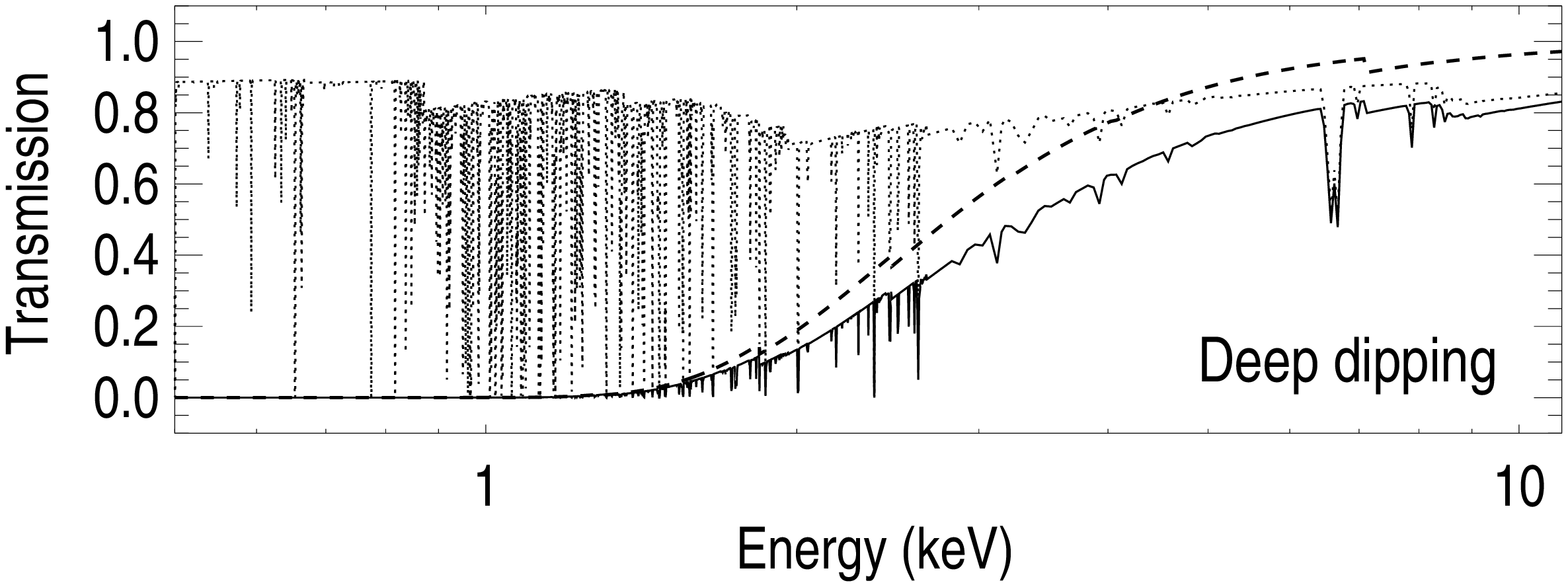}\hfill
\includegraphics[origin=c,angle=-90,width=0.5\textwidth]{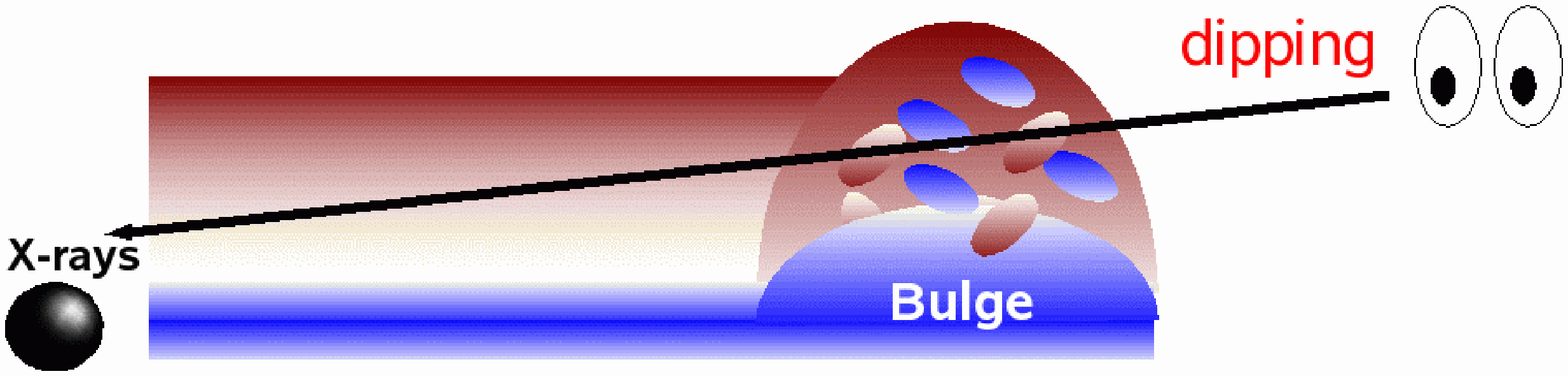}}
{\vspace{-3.4cm}{\bfseries A \hfill B \hfill}}
\caption{{\bfseries A)} Transmission of the ionized (dotted line) and
  neutral absorbers (dashed line), and the total transmission (thick
  line) during persistent ({\bfseries top}) and deep dipping
  ({\bfseries bottom}) intervals of \src\ \citep[adapted
  from][]{1323:boirin05aa}.  During persistent segments, the ionized
  plasma transmits all the photons except those with an energy
  matching the \fetfive\ and \fetsix\ transitions, while during deep
  dipping, the transmission is affected by lines and edges from many
  ions and becomes strongly energy-dependent. The neutral absorption
  is larger during dipping than during persistent states, indicating
  that part of the neutral absorber is located in the binary rather
  than in the interstellar medium, at least during dipping. {\bfseries
  B)} These results suggest that a highly-ionized atmosphere is
  present above the accretion disk and seen in absorption during
  persistent segments ({\bfseries top}).  During dipping ({\bfseries
  bottom}), the bulge passes through the line-of-sight. It's denser, a
  bit less ionized and probably contains clumps of neutral material.}
\label{fig:trans}
\end{figure*}

%
%
\begin{figure*}[t!]
\centerline{\hspace{1.2cm}
\includegraphics[angle=0,width=0.66\textwidth]{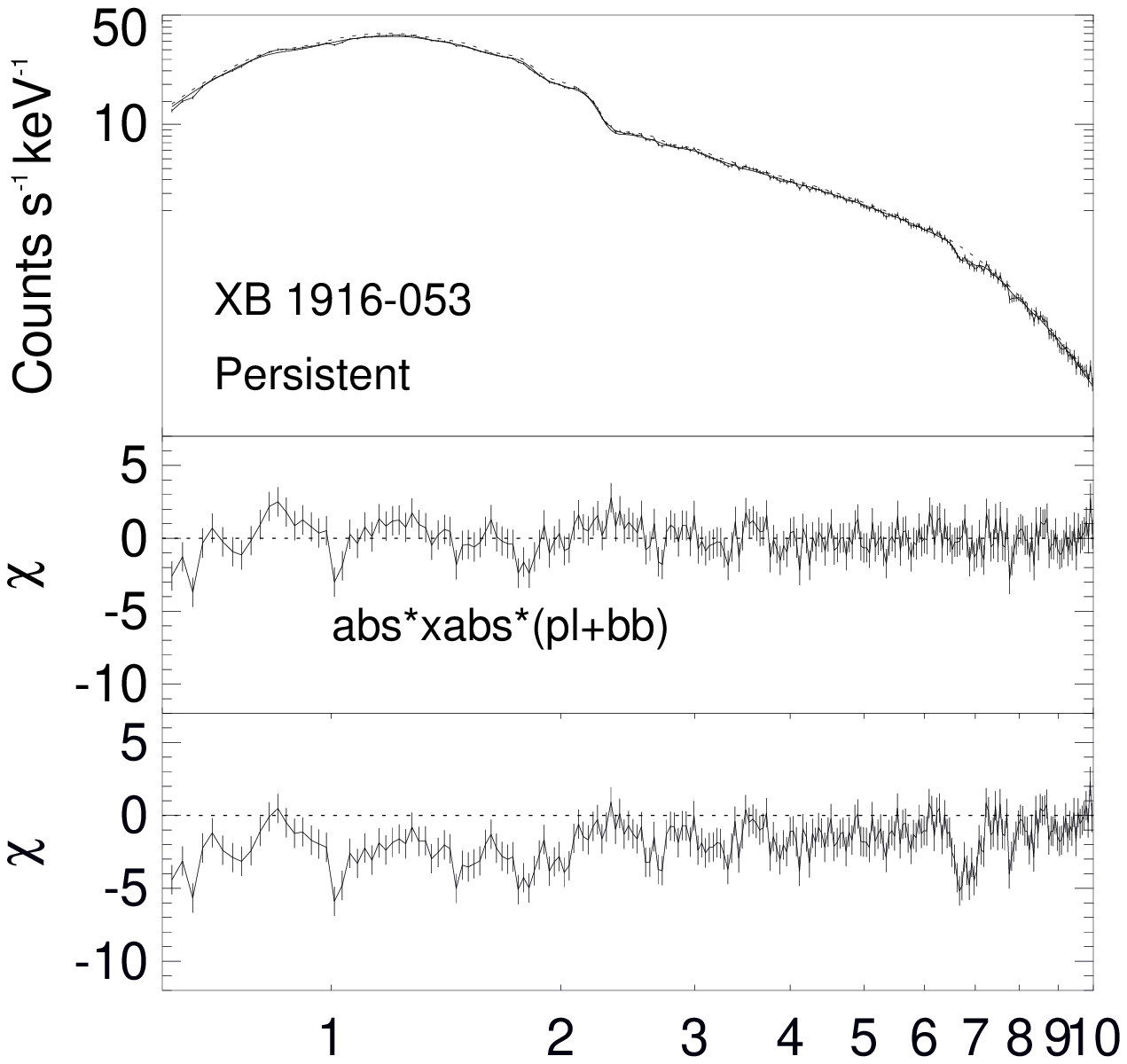}
\hspace{-2.6cm}
\includegraphics[width=0.66\textwidth]{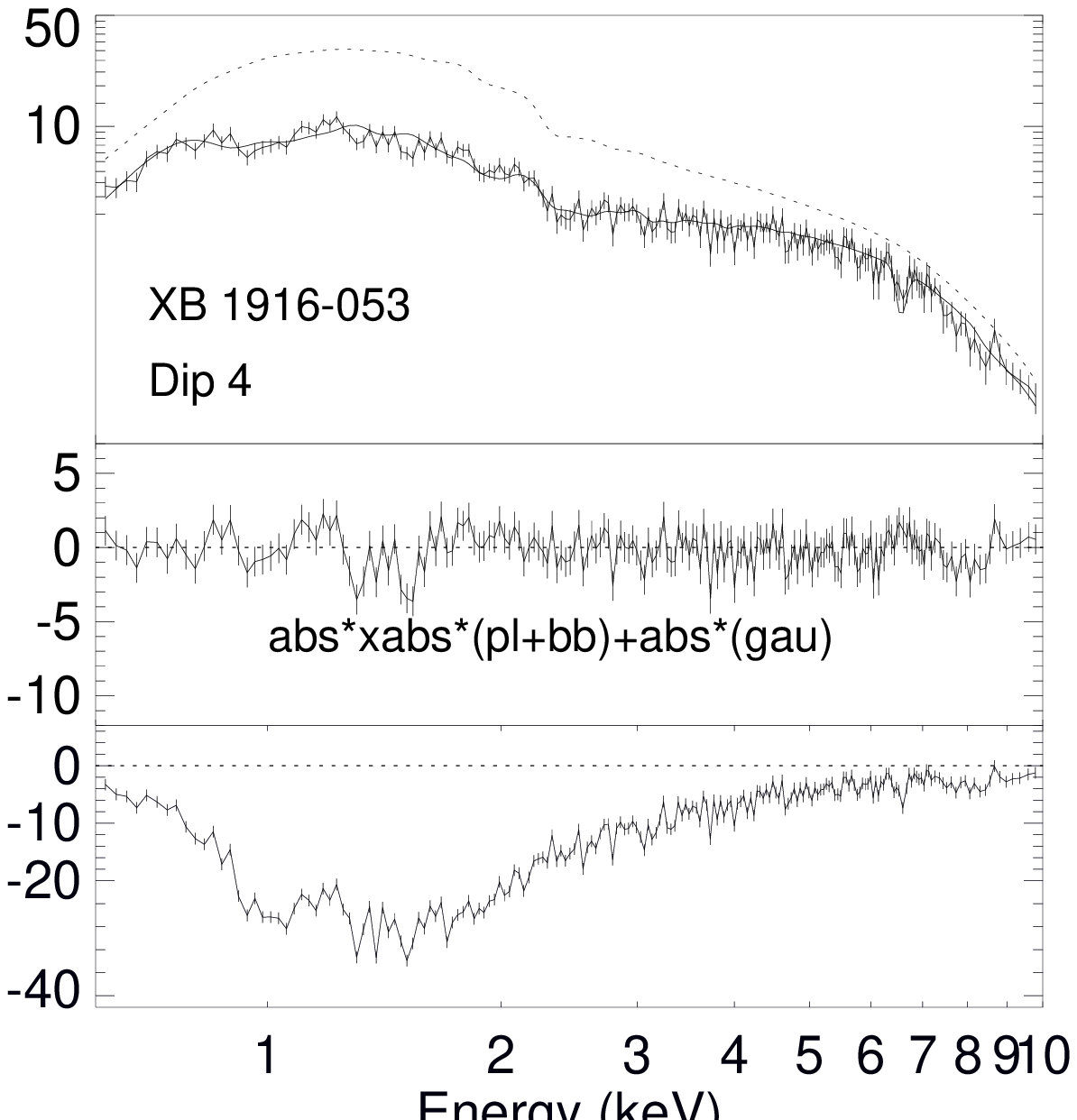}
}
{\vspace{-0.4cm}{\bfseries A \hfill B \hfill}}
\caption{EPIC PN persistent ({\bfseries left}) and dipping ({\bfseries
    right}) spectra of \nineteen\ fit using a photo-ionized absorber
    model \citep[from][]{diaztrigo05aa}. The flat residuals in the
    {\bfseries middle} panel indicate that the fits are good. The
    contribution of the ionized absorber is shown in the {\bfseries
    bottom} panel (by setting \nhxabs\ to 0).  During persistent
    intervals, it produces mainly the narrow absorption lines near
    7~keV, while during dipping, it also produces strong
    energy-dependent absorption throughout the spectrum.}
\label{fig:1916}
\end{figure*}

%
%
\begin{figure*}[b!]
\centerline{\hspace{0.5cm}\includegraphics[width=0.58\textwidth]{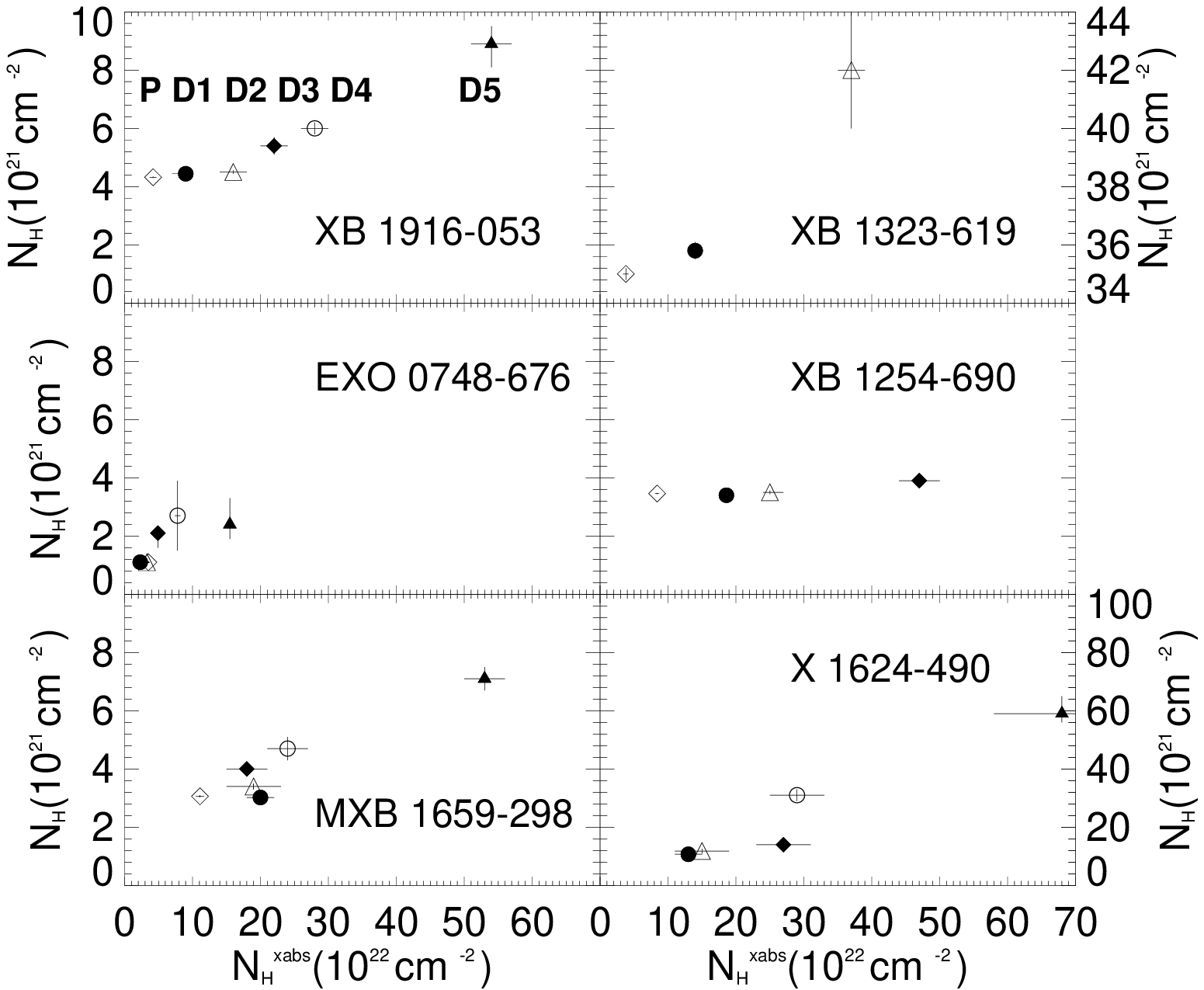}\hspace{-0.9cm}
\includegraphics[width=0.58\textwidth]{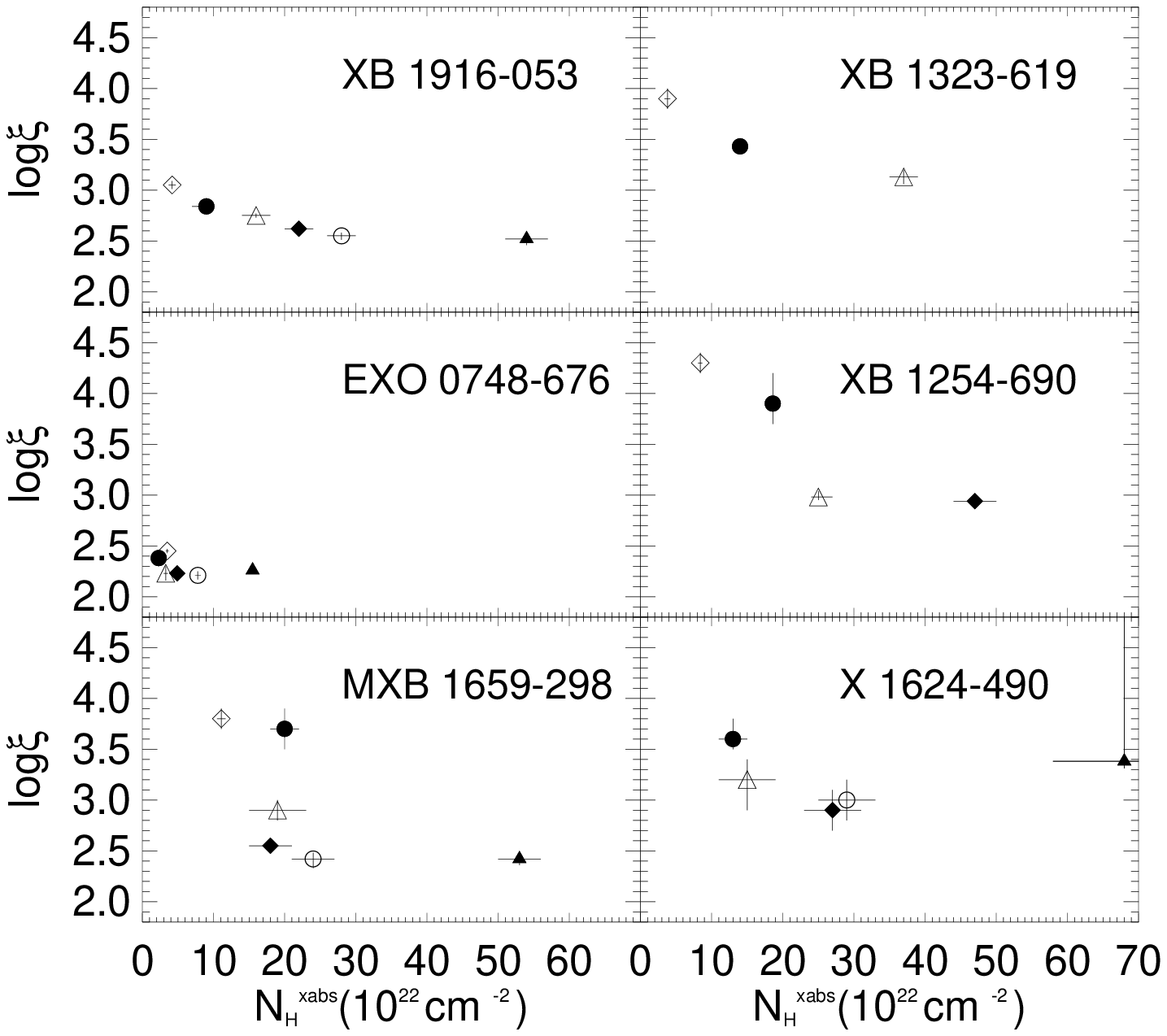}}
{\vspace{-0.4cm}{\bfseries A \hfill B \hfill}}
\caption{ Main properties of the ionized and neutral absorbers in the
  dipping binaries where the photo-ionization absorber model was
  tested \citep[from][]{diaztrigo05aa}. Each symbol represents a
  different intensity stage of the source, from the persistent level,
  P, to the deepest dipping level, D5, as indicated in the top left
  panel.  {\bfseries A)} Column density of the neutral absorber, \nh,
  (including local and interstellar material) versus column density of
  the local ionized absorber, \nhxabs. In all the sources, the amount
  of both neutral and ionized material in the line-of-sight increases
  from persistent to deep dipping stages.  {\bfseries B)} Ionization
  parameter, $\xi$, versus column density, \nhxabs, of the ionized
  absorber.  In all the sources, the ionization parameter decreases
  from persistent to deep dipping while \nhxabs\ increases.}
\label{fig:multi}
\end{figure*}

\end{onecolumn}

\begin{twocolumn}

\section{Results on other dipping sources}

To test the new proposed explanation for the spectral changes,
\citet{diaztrigo05aa} analyzed the EPIC PN data of all the bright
dipping binaries observed by \xmm: \nineteen, \exo, \twelve, \stfs,
\mxb\ and \bigdip\ (see \diaz\ et al. in these proceedings). For each
source (except \stfs\ whose dips were too shallow for the analysis to
be carried out), the persistent and dipping spectra were fit together
with the parameters of the underlying continuum emission tied
together, and the parameters of the absorbers (one neutral, {\tt abs},
and one photo-ionized, {\tt xabs}) left free to vary.  Good fits were
obtained for each source (see the case of \nineteen\ in
Fig.~\ref{fig:1916}).  Thus, the changes in the properties of a
neutral and of an ionized absorber in the line-of-sight can account
for the spectral changes in the continuum and in the narrow features
of all the dipping sources that could be tested so far.  From
persistent to deep dipping, the amount of neutral absorber increases
(Fig.~\ref{fig:multi}~A), corresponding to an increase by a factor
$\sim$2 in the amount of the local material.  At the same time, the
column density of the ionized absorber is found to increase by a
factor of 4 to 12 (Fig.~\ref{fig:multi}~A) while its ionization
parameter decreases (Fig.~\ref{fig:multi}~B).  The changes in this
ionized material clearly play the main role in explaining the overall
energy-dependent spectral changes observed in the dipping sources (see
the bottom panel of Fig.~\ref{fig:1916}~A and B showing the
contribution of the ionized absorber).

\section{Conclusions and prospects}

Modeling the spectral changes between persistent and dipping intervals
is a powerful means of learning about the bulge and the accretion disk
in all the X-ray binaries.  Until now, these spectral changes were
modeled by invoking absorption of a point-like emission region by a
neutral absorber, together with progressive and partial covering of an
extended emission region by another neutral absorber.  We propose a
novel and simpler explanation invoking a neutral absorber and a
photo-ionized absorber.  It was successfully applied to all the bright
dipping sources that could be tested to date: \src, \nineteen, \exo,
\twelve, \mxb\ and \bigdip.  No partial covering was needed,
indicating that none of the underlying X-ray sources requires to be
particularly extended.  The new approach has the strong advantage of
explaining self-consistently the spectral changes both in the
continuum and in the narrow absorption lines that have been revealed
by XMM-Newton.

These results suggest a geometry for X-ray binaries such as drawn in
Fig.~\ref{fig:trans}~B. A highly-ionized plasma is present above the
accretion disk. If the binary is viewed relatively close to edge-on,
the ionized plasma lies in our line-of-sight toward the X-rays emitted
in the vicinity of the compact object, and signatures of the plasma
appear in the spectrum, such as the \fetfive\ and \fetsix\ absorption
lines in the persistent spectrum of \src.  At the azimuth where the
stream of material from the companion star impacts the disk, there is
material projected at higher altitudes above the disk. This bulge or
thickened part of the disk passes through our line-of-sight during
dipping.  Contrary to the complex continuum approach, our modeling of
the dipping spectra indicates that this material is ionized (but less
than the plasma seen during persistent intervals). It probably
contains clumps of neutral material.

The precise distribution of the ionized absorber is unknown.
Possibly, from the surface of the disk to higher altitudes, the
density of the ionized material decreases and hence its ionization
parameter increases.  If the ionized absorber is present at the radius
of the bulge, its layers could be shifted to higher altitudes. This
could explain the differences observed between persistent and dipping
intervals in a given source, and the differences in the absorbers
properties observed from source to source, as a function of
inclination.  In any case, the geometry inferred from the dipping
sources should be valid for all the other accreting binaries which
only differ from the dipping ones in being viewed further away from
the disk plane. This makes the dipping sources among the best targets
to improve our understanding of the disk structure and of the
accretion process.

Here are some of the key issues that we would like to address thanks
to future detailed X-ray observations of the dipping sources:
\begin{itemize}
\item constraints on the distribution of the ionized material: inner
  and outer radii, height, density gradient, ionization gradient,
  composition, velocity (static atmosphere versus out-flowing wind);
\item response of the ionized material to changes in the
  underlying source luminosity or spectral energy distribution;
\item dependence of the properties of the ionized material on the
  system parameters such as the disk size or inclination;
\item role of the reflection (back-scattering) of X-rays onto the
  ionized and neutral materials.
\end{itemize}


\end{twocolumn}

\end{document}
